\documentclass[aps,pra,twocolumn,letterpaper,groupedaddress,10pt]{revtex4-2}
\usepackage{amssymb,amsthm,amsmath,amsfonts}
\usepackage{graphicx,ulem,mathptmx}
\usepackage[shortlabels]{enumitem}
\usepackage[pdftex,dvipsnames,usenames]{xcolor}
\usepackage[colorlinks=true,urlcolor=blue,citecolor=blue,linkcolor=blue]{hyperref}

\begin{document}


\title{Experimental entanglement characterization of two-rebit states}

\author{Nidhin Prasannan}
	\affiliation{Integrated Quantum Optics Group, Institute for Photonic Quantum Systems (PhoQS), Paderborn University, Warburger Stra\ss{}e 100, 33098 Paderborn, Germany}
\author{Syamsundar De}
	\affiliation{Integrated Quantum Optics Group, Institute for Photonic Quantum Systems (PhoQS), Paderborn University, Warburger Stra\ss{}e 100, 33098 Paderborn, Germany}
\author{Sonja Barkhofen}
	\affiliation{Integrated Quantum Optics Group, Institute for Photonic Quantum Systems (PhoQS), Paderborn University, Warburger Stra\ss{}e 100, 33098 Paderborn, Germany}
\author{Benjamin Brecht}
	\affiliation{Integrated Quantum Optics Group, Institute for Photonic Quantum Systems (PhoQS), Paderborn University, Warburger Stra\ss{}e 100, 33098 Paderborn, Germany}
\author{Christine Silberhorn}
	\affiliation{Integrated Quantum Optics Group, Institute for Photonic Quantum Systems (PhoQS), Paderborn University, Warburger Stra\ss{}e 100, 33098 Paderborn, Germany}
\author{Jan Sperling}
    \email{jan.sperling@upb.de}
	\affiliation{Integrated Quantum Optics Group, Institute for Photonic Quantum Systems (PhoQS), Paderborn University, Warburger Stra\ss{}e 100, 33098 Paderborn, Germany}

\begin{abstract}
    We characterize entanglement subject to its definition over real and complex, composite quantum systems.
	In particular, a method is established to assess quantum correlations with respect to a selected number system, illuminating the deeply rooted, yet rarely discussed question of why quantum states are described via complex numbers.
	With our experiment, we then realize two-photon polarization states that are entangled with respect to the notion of two rebits, comprising two two-level systems over real numbers.
	At the same time, the generated states are separable with respect to two complex qubits.
	Among other results, we reconstruct the best approximation of the generated states in terms of a real-valued, local expansion and show that this yields an incomplete description of our data.
	Conversely, the generated states are shown to be fully decomposable in terms of tensor-product states with complex wave functions.
	Thereby, we probe paradigms of quantum physics with modern theoretical tools and experimental platforms that are relevant for applications in quantum information science and technology and connected to the fundamentals of the quantum description of nature.
\end{abstract}

\date{\today}
\maketitle


\section{Introduction}

    The axiomatic formulation of quantum theory \cite{D30,N32} resulted in one of the most successful mathematical descriptions of nature.
    But even quite subtle aspects of this theory defied classical notions, which were thought to be universal, such as locality.
    This resulted in objections---carefully constructed to produce contradictions---that concerned the self-consistency and completeness of quantum theory itself, as famously expressed in the seminal EPR paper \cite{EPR35}.
    Still, even the most abstract debates about the fundamentals of quantum physics led to profound insight and the emergence of previously inconceivable technologies.
    For example, the EPR paradox gave birth to the notion of entanglement \cite{W89,HHHH09} that is recognized today as a key resource for many practical quantum information protocols \cite{NC00,HHHH09}.

    Two core principles of quantum physics are (i) the formulation of this theory in complex Hilbert spaces and (ii) the composition of two quantum systems in terms of tensor-product spaces \cite{D30,N32}.
    The latter constitutes the basis for the joint description of two or more quantum systems, being essential for the notion of entanglement \cite{S35,S36}.
    Less frequently discussed, however, is the first principle which states that wave functions can take complex values, an observation made by Schr\"odinger when introducing his equations of motion \cite{S26}.
    In this context, it is worth reminding ourselves that classical functions---e.g., densities, potentials, temperature distributions, etc.---are typically based on real values, and the introduction of complex numbers to describe the state of a quantum system was likely not a canonical choice.
    See also the recent contributions \cite{K20,RTWTTGAN21} in this context, theoretically discussing that real Hilbert spaces would lead to measurable contradictions.

    In their work \cite{CFR01}, Caves, Fuchs, and Rungta combined those two seemingly harmless requirements of quantum theory, (i) complex numbers and (ii) tensor products, to theoretically predict the existence of a state that is nonentangled (i.e., separable \cite{W89}) when using complex Hilbert spaces but entangled when restricting quantum physics to real numbers.
    This Caves-Fuchs-Rungta (CFR) state reads
    \begin{equation}
        \label{eq:CFRstateOriginal}
        \hat\rho=\frac{1}{4}\left(\hat\sigma_0\otimes\hat\sigma_0
        +r\hat\sigma_y\otimes\hat\sigma_y\right),
    \end{equation}
    where $-1\leq r\leq 1$ and $r\neq 0$ and using the Pauli matrix $\hat\sigma_y$ and the identity $\hat\sigma_0$.
    The CFR state is defined over two two-dimensional, real Hilbert spaces.
    Analogously to the notion of a qubit---the basic unit of quantum information that is encoded in a two-dimensional, complex Hilbert space---, the concept of a rebit---a quantum bit in real spaces---was thereby introduced, too.
    Despite the intriguing property of being entangled (over reals) and separable (in complex spaces) at the same time \cite{CFR01}, which is discussed later in detail, this state, which has been theoretically proposed two decades ago, has not been experimentally analyzed regarding its entanglement properties to our best knowledge.

    Exceeding fundamental questions, real Hilbert spaces, specifically rebits, have been found to offer a versatile resource for quantum information processing.
    For instance, rebit-based forms of universal quantum computing were developed \cite{RG02,ABW13}. 
	And quantum simulators were shown to benefit from employing real Hilbert spaces as well \cite{MMG09,KNY18}.
    Furthermore, the relation of rebit entanglement to the states' purity was studied \cite{BPCP02}, where the purity itself can be considered as a useful quantum resource \cite{SKWGB18}.
    In general, the quantum resources that are connected to imaginary parts of the density operator can be quantified rigorously \cite{WKRSXLGS20}.
    Moreover, while qubits can only share maximal entanglement in pairs, this limitation holds, surprisingly, no longer true for entanglement based on real numbers \cite{W12}.
    In fact, an arbitrary number of rebits can be maximally entangled \cite{W12}, being a quite remarkable feature from a quantum communication perspective.
	Comprehensive investigations further revealed that one can embed real, complex, and even quaternionic quantum theories into a joint framework \cite{BPCP03,B12}, being relevant for dynamic \cite{B12} and thermodynamic considerations \cite{ABW13}.
    Using such methods, it was also argued via quantum-dynamical group representations and their symmetries that, among the different theories based on distinct sets of numbers, the complex version necessarily emerges as the preferred one \cite{MO17,CL17}.

    Because of the fundamental and application-based importance of entanglement in real Hilbert spaces, criteria to detect and quantify the corresponding kind of entanglement have been developed \cite{CFR01}.
	This includes uncertainty relations for rebit entanglement \cite{A19} and negativities of distributions that resemble phase-space quasiprobabilities \cite{DGBR15,SW18}.
	In addition, measurement and reconstruction approaches have been theoretically studied in this context \cite{H17,H19}.
    Still, despite the progress in detecting rebit entanglement, experiments which complement such theoretical works are missing to date.

    In this contribution, we experimentally realize CFR states and devise and apply a characterization framework to access entanglement over complex and real Hilbert spaces.
    Our theoretical tools includes a construction of entanglement witnesses and the representation of entanglement in terms of so-called entanglement quasiprobabilities and a remaining, locally nondecomposable element.
    Our experiment uses a state-of-the-art single-photon source with a computational basis in the horizontal and vertical polarization.
    By producing mixtures of circularly polarized light, we are then able to implement rebit states with high similarities to the targeted CFR states.
    Our witnesses reveal a close-to-maximal rebit entanglement.
    Conversely, the entanglement quasiprobabilities yield a separable decomposition of the produced state in terms of tensor-products of complex qubits.
    Furthermore, it is shown that a rebit description of the state is incomplete with an almost maximally nondecomposable contribution, constituting the origin of the rebit-only entanglement under study.
    Therefore, we experimentally implement and characterize a unique form of quantum correlation that depends on the choice of numbers.

\section{Theory}

    Suppose $\{|0\rangle,\ldots,|d-1\rangle\}$ is the computational basis for Alice's ($d=d_A$) and Bob's ($d=d_B$) subsystem.
    In our experiment, we have $d_A=d_B=2$, with $|H\rangle=|1\rangle$ and $|V\rangle=|0\rangle$ for horizontal and vertical polarization, respectively.
    A separable state over a real Hilbert spaces is defined in terms of a classical joint probability $P(a,b)$, where $P(a,b)\geq0$ and $\sum_{a,b}P(a,b)=1$, and tensor-product states, $|a\rangle\otimes|b\rangle=|a,b\rangle=|ab\rangle$ with $|a\rangle\in\mathbb R^{d_A}$ and $|b\rangle\in\mathbb R^{d_B}$, as
    \begin{eqnarray}
        \label{eq:DefSep}
        \hat\rho_{\text{$\mathbb R$-sep.}}=\sum_{a,b}P(a,b)|a,b\rangle\langle a,b|,
    \end{eqnarray}
    analogously to complex Hilbert spaces \cite{W89}.
    Whenever the state cannot be written in this form, it is inseparable, i.e., entangled, in the bipartite scenario studied here.
    The CFR state in Eq. \eqref{eq:CFRstateOriginal}, which can be also expressed as
	\begin{equation}
	\label{eq:CFRstate}
	\begin{aligned}
		\hat\rho=&\frac{q}{2}\left(
			|L,L\rangle\langle L,L|
			+|R,R\rangle\langle R,R|
		\right)
		\\
		&+\frac{1-q}{2}\left(
			|L,R\rangle\langle L,R|
			+|R,L\rangle\langle R,L|
		\right)
		\\
		=&\frac{1}{4}\begin{pmatrix}
			1 & 0 & 0 & 1-2q
			\\
			0 & 1 & 2q-1 & 0
			\\
			0 & 2q-1 & 1 & 0
			\\
			1-2q & 0 & 0 & 1
		\end{pmatrix},
	\end{aligned}
	\end{equation}
	where $0\leq q=(r+1)/2\leq 1$, is a two-rebit mixed state with a real-valued density matrix in the computational basis.
	Particularly, it is a statistical mixture of complex tensor-product states with circular polarization, $|R\rangle=(|1\rangle+i|0\rangle)/\sqrt{2}$ and $|L\rangle=(|1\rangle-i|0\rangle)/\sqrt{2}$, thus clearly $\mathbb C$ separable.
	The task now is to find applicable criteria to prove $\mathbb R$ entanglement.
    
    For detecting $\mathbb C$ entanglement, measurable entanglement criteria are indeed available, resulting in the notion of entanglement witnesses \cite{HHH05,T05}; see also the review \cite{GT09}.
    Applying the same optimization approach as established in Ref. \cite{SV09} for complex numbers, we can directly formulate separability constraints for observables $\hat L$ in real Hilbert spaces as
    \begin{eqnarray}
        \label{eq:EntCrit}
        g^{(\mathbb R)}_{\min}\leq \langle\hat L\rangle_{\text{$\mathbb R$-sep.}}=\mathrm{tr}(\hat\rho_{\text{$\mathbb R$-sep.}}\hat L)\leq g^{(\mathbb R)}_{\max},
    \end{eqnarray}
    where $g^{(\mathbb R)}_{\min}$ and $g^{(\mathbb R)}_{\max}$ are the maximal and minimal expectation value for $\mathbb R$-separable states.
    The latter bounds can be obtained by solving so-called separability eigenvalue equations,
    \begin{equation}
        \label{eq:SEP}
        \hat L_a|b\rangle=g^{(\mathbb R)}|b\rangle
        \text{ and }
        \hat L_b|a\rangle=g^{(\mathbb R)}|a\rangle,
    \end{equation}
    with $\hat L_a=(\langle a|{\otimes}\hat 1_B)\hat L(|a\rangle{\otimes}\hat 1_B)$ and $\hat L_b=(\hat 1_A{\otimes}\langle b|)\hat L(\hat 1_A{\otimes}|b\rangle)$, providing the sought-after bounds $g^{(\mathbb R)}_{\min}=\min g^{(\mathbb R)}$ and $g^{(\mathbb R)}_{\max}=\max g^{(\mathbb R)}$.
    See the Supplemental Material (SM) for a derivation \cite{SuppMat}.
    If the expectation value $\langle \hat L\rangle$ exceeds either bound in Eq. \eqref{eq:EntCrit}, $\mathbb R$ entanglement is certified.
    Similarly to the ordinary eigenvalue problem that yields optimal expectation values, our $\mathbb R$-separability bounds are determined through the maximal and minimal separability eigenvalues from Eq. \eqref{eq:SEP}.
    See Ref. \cite{SV09} for detailed discussions for the related $\mathbb C$-entanglement witnesses construction and, for example, Refs. \cite{GPMMSVP14,GSVCRTF15} for their experimental applications.

    For instance, this approach can be used to obtain Clauser-Horne-Shimony-Holt-type inequalities \cite{CHSH69} when choosing observables like $\hat L=L_z\hat\sigma_z\otimes\hat\sigma_z+L_x\hat\sigma_x\otimes\hat\sigma_x$ \cite{T05}.
    (See, e.g., Ref. \cite{RTWTTGAN21} for a recent Bell-type inequality as well.)
    Then, Eq. \eqref{eq:SEP} can be easily solved (see the SM \cite{SuppMat}) and results in the separable bounds $g^{(\mathbb R)}_{\max}=\max\{|L_z|,|L_x|\}$ and $g^{(\mathbb R)}_{\min}=-g^{(\mathbb R)}_{\max}$ that correspond to separability eigenvectors $|a,b\rangle$ in which the local components are in horizontal, vertical, diagonal [$|D\rangle=(|0\rangle+|1\rangle)/\sqrt{2}$], and antidiagonal [$|A\rangle=(|0\rangle-|1\rangle)/\sqrt{2}$] polarization, which are all in $\mathbb R^2$.
    Bell states can maximally violate these bounds, and the corresponding maximal and minimal (ordinary) eigenvalues are $g_{\max}=|L_z|+|L_x|=-g_{\min}$, exceeding their separable counterparts.
    However, this choice of $\hat L$ does not make a distinction between $\mathbb C$ entanglement and $\mathbb R$ entanglement because $g^{(\mathbb R)}_{\min}=g^{(\mathbb C)}_{\min}$ and $g^{(\mathbb R)}_{\max}=g^{(\mathbb C)}_{\max}$ hold true.

    A more viable choice would be an even simpler observable, $\hat L=\hat\sigma_y\otimes\hat\sigma_y$, which has a real-valued matrix representation.
    The ordinary eigenvalue equations, complex separability eigenvalue equations \cite{SV09}, and real separability eigenvalue equations [Eq. \eqref{eq:SEP}] can be also straightforwardly solved \cite{SuppMat}.
    This yields $g_{\max}=g_{\max}^{(\mathbb C)}=1$ and $g_{\min}^{(\mathbb C)}=g_{\min}=-1$ as bounds to the expectation values $\langle\hat L\rangle$ of general and $\mathbb C$-separable states, thus not allowing for witnessing $\mathbb C$ entanglement.
    Note that these bounds are obtained for products of circularly polarized states, $|L\rangle$ and $|R\rangle$, with a complex expansion in the computational basis.

    In contrast, for $\mathbb R$-separable states, the reduced operators (e.g., $\hat L_a$) in Eq. \eqref{eq:SEP} vanish since we find that $\hat L_a=(\langle a|{\otimes}\hat\sigma_0 )\hat\sigma_y\otimes\hat\sigma_y(|a\rangle{\otimes}\hat\sigma_0)=\langle a|\hat\sigma_y|a\rangle\hat\sigma_y=0\,\hat\sigma_y$ holds true for all $|a\rangle\in\mathbb R^2$.
    This yields $g^{(\mathbb R)}=0$, thus $g_{\min}^{(\mathbb R)}=0=g_{\max}^{(\mathbb R)}$.
    Consequently, whenever
    \begin{equation}
        \label{eq:SigmaYCriterion}
        \langle\hat\sigma_y\otimes\hat\sigma_y\rangle\neq0
    \end{equation}
    is measured, $\mathbb R$ entanglement is verified, but $\mathbb C$ entanglement is not.
    For example, for the CFR states in Eq. \eqref{eq:CFRstateOriginal}, we expect $\langle\hat\sigma_y\otimes\hat\sigma_y\rangle=r$, which constitutes a maximal violation of the $\mathbb R$-separable bound zero for $r=\pm1$---likewise, $q\in\{1,0\}$ for Eq. \eqref{eq:CFRstate}.  
    It is worth outlining that, for higher-dimensional systems ($d_A,d_B>2$), any $\hat L$ that is a linear combination of tensor products of skew-symmetric operators (i.e., $\hat \alpha\otimes \hat \beta$, with $\hat\alpha=-\hat\alpha^\mathrm{T}\in\mathbb R^{d_A\times d_A}$ and $\hat\beta=-\hat\beta^\mathrm{T}\in\mathbb R^{d_B\times d_B}$) can be applied in a similar manner.

    In summary, we introduced a generally applicable method to construct measurable witnesses for $\mathbb R$ entanglement [cf. Eqs. \eqref{eq:EntCrit} and \eqref{eq:SEP}] and to tell such quantum correlations from $\mathbb C$-entangled states apart.
    Moreover, this approach can be straightforwardly generalized to multipartite systems, too, as done for complex Hilbert spaces in Ref. \cite{SV13}.

    Furthermore, it was shown that any $\mathbb C$-separable state and $\mathbb C$-inseparable state too can be fully expanded in terms of a pseudomixture of tensor-product states when allowing for $P(a,b)<0$ in the definition of separability \cite{STV98}, defining the notion of entanglement quasiprobabilities \cite{SV09quasi} that have been recently experimentally reconstructed for Bell states \cite{SMBBS19}.
    Interestingly, the construction of such quasiprobabilities is also based on equations of the form of Eq. \eqref{eq:SEP} when replacing $\hat L$ with the density operator $\hat\rho$, which also allows one to find the best separable decomposition \cite{LS98,KL01}.
    For real Hilbert spaces, where $\hat\rho=\mathrm{Re}(\hat\rho)$, the quasiprobability method was derived in Ref. \cite{SW18}.
    However, already for two rebits, a full decomposition of general $\mathbb R$-inseparable states $\hat\rho$ is not possible in such a generalized local form.
    Rather, a residual component can exist that cannot be decomposed in terms of any linear combination of tensor products of rebit states \cite{SW18}; specifically, we have
    \begin{equation}
        \label{eq:Residuum}
        \hat\rho=\sum_{a,b}P(a,b)|a,b\rangle\langle a,b|+\rho_{\text{res.}}\hat\sigma_y\otimes\hat\sigma_y,
    \end{equation}
    proving the incompleteness of a local---even when including quasiprobabilities---description of $\mathbb R$-inseparable rebit states.

    Local decompositions can be reconstructed from data using the approach in Ref. \cite{SMBBS19}.
    This algorithm is based on the observation that general two-qubit and two-rebit states $\hat\rho$ can be recast into the so-called standard form \cite{LMO06},
    \begin{equation}
        \label{eq:StdForm}
        \hat\rho_\text{std.}=
            \rho_0\hat\sigma_0\otimes\hat\sigma_0
            +\rho_z\hat\sigma_z\otimes\hat\sigma_z
            +\rho_x\hat\sigma_x\otimes\hat\sigma_x
            +\rho_y\hat\sigma_y\otimes\hat\sigma_y,
    \end{equation}
    which is diagonal as an expansion in terms of tensor products of Pauli operators.
    This is achieved through local operations,
    \begin{eqnarray}
        \label{eq:LocalInvertible}
        \hat\rho=\frac{
            (\hat A\otimes\hat B)\hat\rho_\text{std.}(\hat A\otimes\hat B)^\dag
        }{\mathrm{tr}\left[
            (\hat A\otimes\hat B)\hat\rho_\text{std.}(\hat A\otimes\hat B)^\dag
        \right]},
    \end{eqnarray}
    thus not affecting separability.
    Importantly, for real and complex numbers, the local operations $\hat A\otimes\hat B$ are also real and complex maps, respectively.
    This follows from diagonalization techniques when computing the standard form from data;
    see the SM \cite{SuppMat} for technical details.

    For the CFR state in Eq. \eqref{eq:CFRstateOriginal}, which is already in standard form, we expect a $\mathbb C$-separable decomposition as given in Eq. \eqref{eq:CFRstate}.
    For a local decomposition over $\mathbb R$, however, the analytical formula in Ref. \cite{SW18} predicts local states $|a,b\rangle\in\{|HH\rangle,|HV\rangle,|VH\rangle,|VV\rangle,|DD\rangle,|DA\rangle,|AD\rangle,|AA\rangle\}$ with probabilities $P(a,b)=1/8$, which only resolve the (normalized) identity part of the CFR state, $\hat\sigma_0\otimes\hat\sigma_0/4$.
    And the nondecomposable, residual component is $r\hat\sigma_y\otimes\hat\sigma_y/4$ [i.e., $\rho_\mathrm{res.}=r/4$ in Eq. \eqref{eq:Residuum}].
    In general, the best local approximation $\hat\rho'=\sum_{a,b}P(a,b)|a,b\rangle\langle a,b|$, including cases $P(a,b)\geq0$ and $P(a,b)<0$, of two rebit states in standard form, Eq. \eqref{eq:StdForm}, has a Hilbert-Schmidt distance to the full density operator that is given by \cite{SW18}
    \begin{equation}
        \|\hat\rho-\hat\rho'\|=2|\rho_y|.
    \end{equation}
    Since physical states obey $|\rho_y|\leq 1/4$, we have a maximal difference of $1/2$.
    In fact, this value is expected for CFR states with $q\in\{0,1\}$ (likewise, $r=\pm1$).

    The ability to reconstruct best local decomposition of the state is relevant for our experimental application because it allows us to quantify the degree of incompleteness of an $\mathbb R$-local description as described above, in addition to the aforementioned witnessing of $\mathbb R$ entanglement.
    It is further worth recalling that one expects a zero distance for complex Hilbert spaces since all states are $\mathbb C$ local, i.e., fully decomposable as a pseudomixture of tensor-product states \cite{STV98}.

\section{Experiment}

\begin{figure}[t]
    \includegraphics[width=\columnwidth]{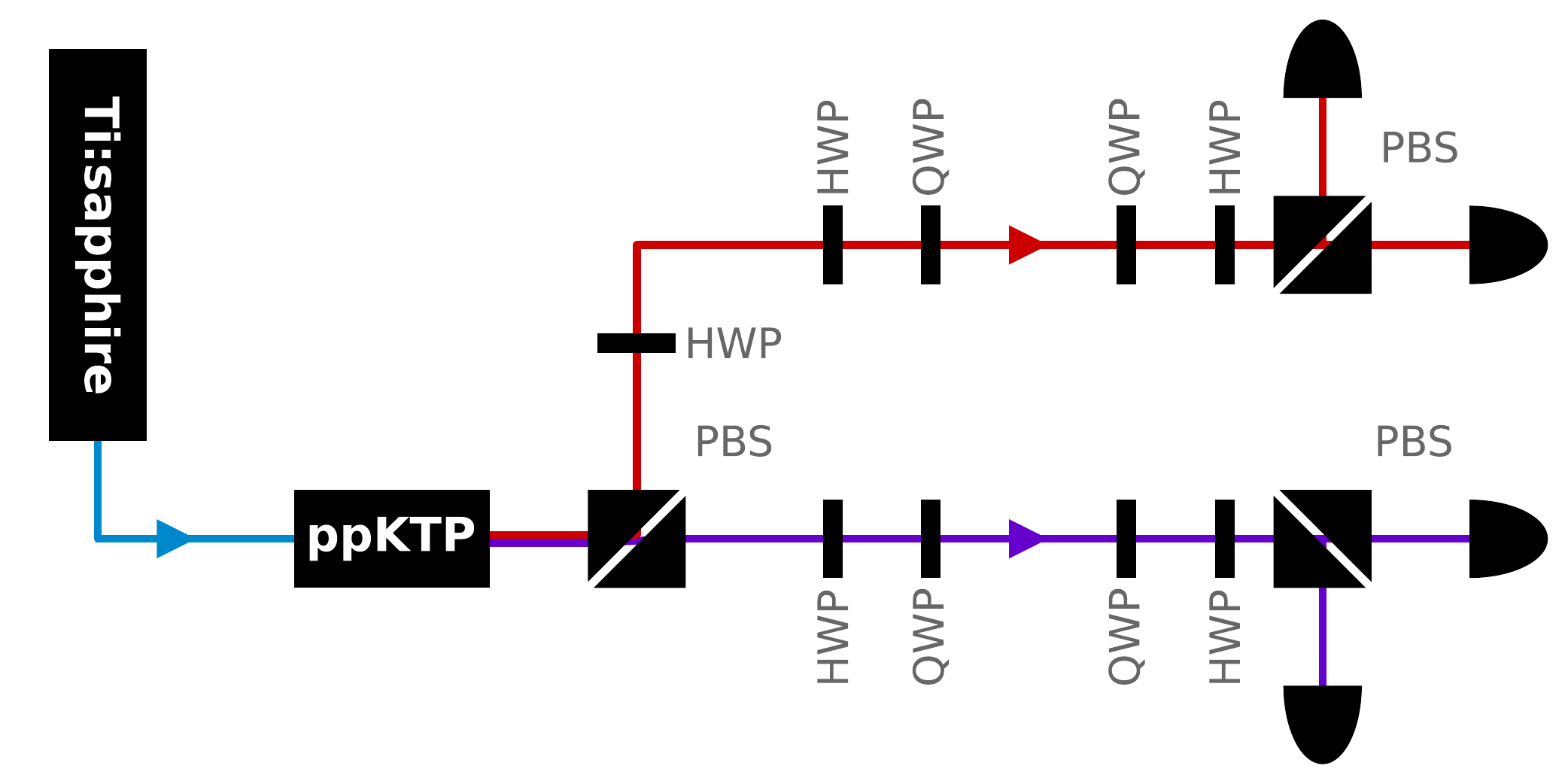}
    \caption{
        Setup outline.
        A laser (Ti:sapphire) pumps a photon-pair source (ppKTP).
        Signal and idler photons are separated with a polarizing beam splitter (PBS).
        Combinations of half- and quarter-wave plates (HWP and QWP, respectively) allow for polarization rotations.
        The first HWP-QWP rotation prepares circularly polarized light in both arms.
        The second QWP-HWP rotation, together with PBSs and single-photon counters, allows for a joint polarization analysis of Alice's and Bob's photons.
    }\label{fig:setup}
\end{figure}

    To experimentally demonstrate two-rebit entanglement (see Fig. \ref{fig:setup} for the experimental layout), we use a highly efficient waveguide-based parametric down-conversion source \cite{HABDMS13,MPEQDBS18}, producing a $|VH\rangle$ photon pair.
    For this purpose, a type-II phase-matched periodically poled potassium titanyl phosphate waveguide is pumped by a femtosecond laser pulse with a central wavelength of $770\,\mathrm{nm}$.
    The two generated, down-converted photons at $1540\,\mathrm{nm}$ telecom wavelength are separated by a polarization beam splitter (PBS).
    Vertically polarized photons are then rotated to horizontal polarization, which yields horizontally polarized photons in both arms, $|HH\rangle$.
    A first combination of one half-wave plate (HWP) and one quarter-wave plate (QWP) can be used in both arms to access all kinds of polarization product states.
    Here, by setting the angles of QWPs at $\pm45^\circ$ (HWPs are at $0^\circ$), we produced all four combinations of left- and right-circularly polarized photon pairs, $\{|LL\rangle,|LR\rangle,|RL\rangle,|RR\rangle\}$.

    After that, second combinations of HWPs and QWPs, together with PBSs, allows for a full state polarization tomography \cite{AJK05}.
    This polarization-resolved measurement with four superconducting nanowire single-photon detectors provides access to all thirty-six coincidence probabilities in the $z$, $x$, and $y$ bases for both Alice's photon and Bob's photon.
	We recorded for each of the four produced polarization states and each of the three detector polarization settings per subsystem $\approx100\,000$ events, resulting in ca. $3.6$ million data points, demonstrating the high stability of our setup.
	Combining the data in which both photons have the same circular polarization and those with different circular polarizations leaves us with a randomization that realizes the CFR states for the parameters $q=1$ and $q=0$ in Eq. \eqref{eq:CFRstate}, respectively.
	Equivalently, one could directly produce CFR states by randomly changing the wave plates, resulting in the same mixture though.
	Another proposal exists to generate mixed states in which mixing is realized by tracing over other degrees of freedom \cite{WABGJJKMP05}.

\section{Results}

\begin{figure}[t]
	\includegraphics[width=.4\columnwidth]{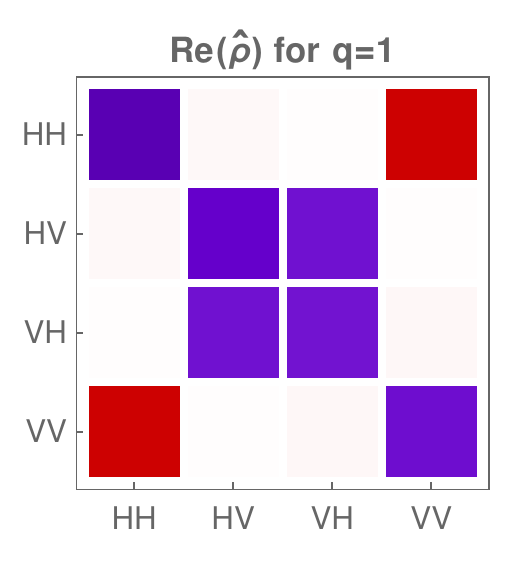}
	\includegraphics[width=.4\columnwidth]{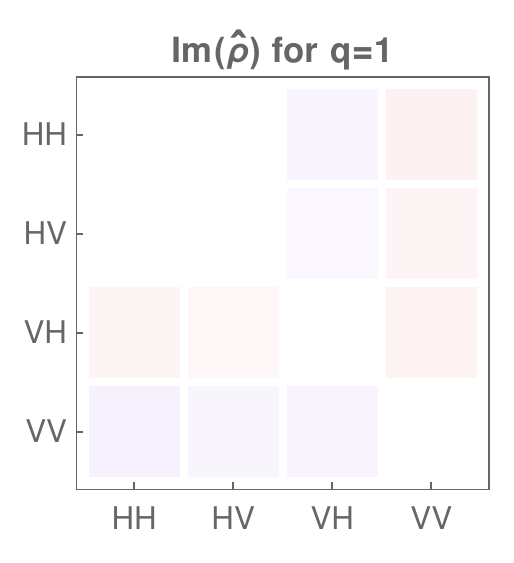}
	\includegraphics[width=.125\columnwidth]{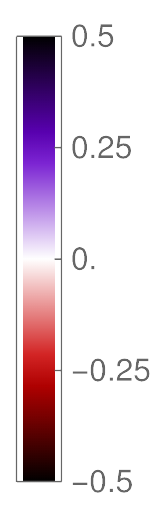}
	\\
	\includegraphics[width=.4\columnwidth]{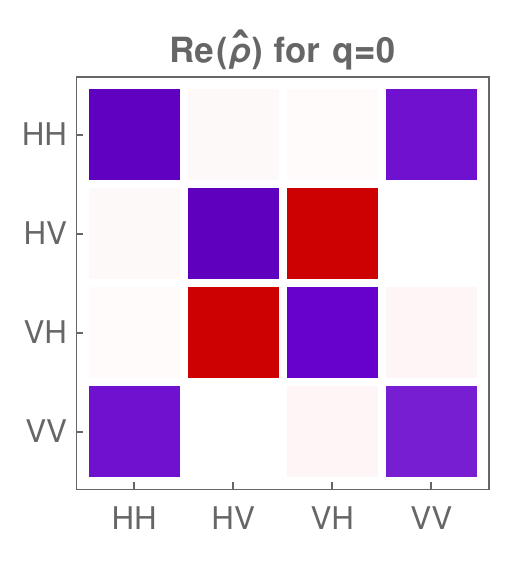}
	\includegraphics[width=.4\columnwidth]{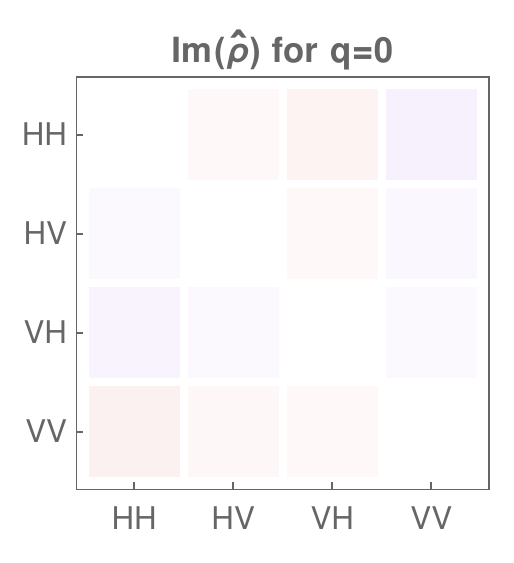}
	\includegraphics[width=.125\columnwidth]{reconstructed_stateLegend.pdf}
	\caption{
		Real (left) and imaginary (right) part of reconstructed density matrix of CFR states in Eq. \eqref{eq:CFRstate} for $q=1$ (top) and $q=0$ (bottom).
		The imaginary part is close to zero, thus resembling the desired two-rebit state.
		In fact, the similarities with the targeted CFR states are $(99.639\pm0.011)\%$ and $(99.667\pm0.011)\%$ for the top and bottom state, respectively.
		The absolute uncertainty for all depicted matrix elements is below $0.0008$ (see the SM \cite{SuppMat} for details).
	}\label{fig:RecDensity}
\end{figure}

	Figure \ref{fig:RecDensity} shows the reconstructed real and imaginary parts of the two produced rebit states.
	For both states, the contribution of the imaginary part is negligible.
	Specifically, the largest amplitude of an imaginary part is for both states about 20 times smaller than for the real part.
	Furthermore, the overlap with the targeted two-rebit states exceeds $99\%$, proving the reliability of the realized state generation.
	Similarities of states $\hat\rho$ and $\hat\rho^\prime$ are quantified through the correlation coefficient $S=\mathrm{tr}(\hat\rho\hat\rho^\prime)/[\mathrm{tr}(\hat\rho^2)\mathrm{tr}(\hat\rho^{\prime 2})]^{1/2}$, which is directly accessible through our measurement and one for $\hat\rho=\hat\rho^\prime$. 

\begin{figure*}
    \includegraphics[width=.24\textwidth]{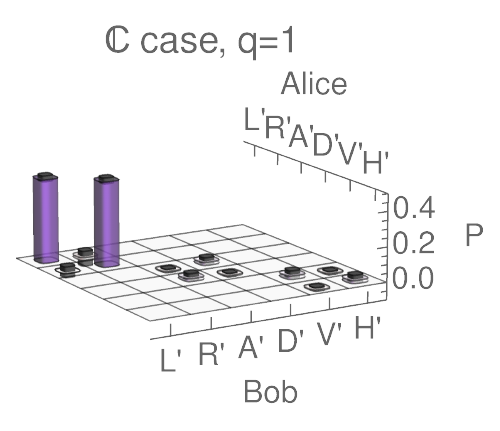}
    \includegraphics[width=.24\textwidth]{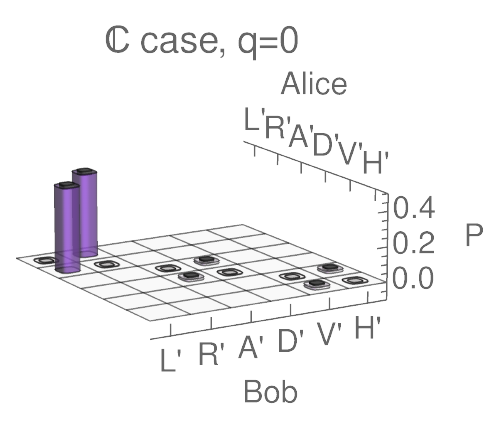}
    \includegraphics[width=.24\textwidth]{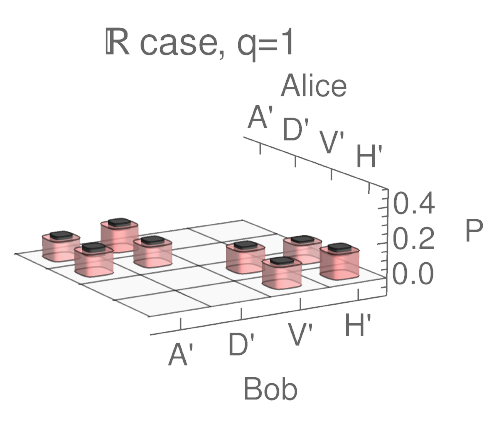}
    \includegraphics[width=.24\textwidth]{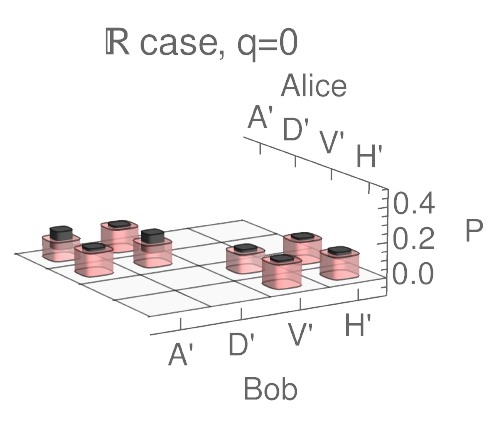}
    \caption{%
        Reconstructed distributions $P(a',b')$ for the best local decomposition in terms of tensor-product states $|a',b'\rangle$ for Alice and Bob; see text for details.
        Plots correspond to CFR states with $q=1$ and $q=0$ for complex and real Hilbert spaces, as indicated by labels.
        For an improved visibility, a five standard-deviation error margin is depicted as black bars \cite{SuppMat}.
    }\label{fig:QuasiPexp}
\end{figure*}

	We can now apply the rebit entanglement criterion in Eq. \eqref{eq:SigmaYCriterion}.
	For the experimentally realized states, we then find a highly significant certification of rebit entanglement,
	\begin{equation}
	\label{eq:WitnessResult}
	\begin{aligned}
		0\neq\langle\hat\sigma_y\otimes\hat\sigma_y\rangle
		=\left\lbrace\begin{array}{rcl}
			0.9634\pm0.0006 & \text{for} & q=1,
			\\
			-0.9631\pm0.0006 & \text{for} & q=0,
		\end{array}\right.
	\end{aligned}
	\end{equation}
	being close to the maximal violations, $\pm1$.
	Thus, $\mathbb R$ entanglement has been experimentally verified.

	What is left is demonstrating that the CFR states realized do not exhibit bipartite qubit entanglement.
	In addition, we are going to show that the rebit entanglement does indeed originate from the nondecomposability of the produced state in terms of local ones.

	For this purpose, we decomposed the state in terms of separable states for the complex and real scenarios; see Fig. \ref{fig:QuasiPexp}.
    It is worth mentioning that Eq. \eqref{eq:LocalInvertible} transforms tensor-product states in a local manner, $|a'\rangle\propto\hat A|a\rangle$ and $|b'\rangle\propto\hat B|b\rangle$, which are used in the plot.
    The complex cases in Fig. \ref{fig:QuasiPexp} describe the states as a balanced and nonnegative mixture of two states with circular polarization for Alice and Bob [Eq. \eqref{eq:CFRstate}], up to negligible contributions from other polarizations.
	By comparison, the best local approximation in the two-rebit case is a uniform mixture of diagonal and antidiagonal as well as horizontal and vertical basis states, which confirms the theoretical predictions.

    To assess how well the best local approximations found describe the states produced, one can determine the distance of the directly reconstructed state $\hat\rho$ (Fig. \ref{fig:RecDensity}) and the density operators $\hat\rho'=\sum_{a',b'}P(a',b')|a',b'\rangle\langle a',b'|$ of the best separable approximations according to Fig. \ref{fig:QuasiPexp}.
	We obtain the distances
	\begin{eqnarray}
	    \|\hat\rho-\hat\rho'\|=\left\lbrace
	    \begin{array}{rcl}
	        	0.000\pm 0.004 & \text{for} & q=1,
	        	\\
	        	0.000\pm 0.004 & \text{for} & q=0,
	    \end{array}
	    \right.
	\end{eqnarray}
	for the complex case.
	This value is zero within the numerical precision used ($10^{-9}$), showing that the complex state is fully described through linear combinations of tensor products of local states.
	By contrast, the real-valued scenario yields
	\begin{eqnarray}
	    \|\hat\rho-\hat\rho'\|=\left\lbrace
	    \begin{array}{rcl}
	        	0.483090\pm 8\times10^{-6} & \text{for} & q=1,
	        	\\
	        	0.482749\pm 8\times10^{-6} & \text{for} & q=0,
	    \end{array}
	    \right.
	\end{eqnarray}
	which is close to the maximal value $1/2$ as the result of a significant nondecomposable, residual component [cf. Eqs. \eqref{eq:Residuum} and \eqref{eq:WitnessResult}].
	This experimentally demonstrates that tensor products of states over real numbers are incomplete for the decomposition of CFR states. 

    Therefore, we successfully characterized a two-rebit state that is a statistical mixture of product states over complex Hilbert spaces but inseparable and locally nondecomposable for real Hilbert spaces.
    This was achieved by preparing mixtures of two-photon states with circular polarizations that results in a density matrix with vanishing imaginary parts in the computational basis of horizontal and vertical polarization.
    The witnessing shows close to maximal violations of a separability constraint for real vector spaces.
    Conversely, in terms of complex spaces, a nonnegative decomposition of the generated state in terms of tensor-products is achieved, demonstrating two-qubit separability.

\section{Conclusion}

    We introduced and applied theoretical methods to characterize and compare entanglement that is defined over real and complex Hilbert spaces.
    By realizing two-rebit states, we addressed the open problem of an experimental characterization of an interesting form of entanglement.
    That is, we demonstrated that the produced CFR states are separable and inseparable at the same time, depending on the choice of the number system.
    Moreover, we reconstructed the actual decomposition in terms of tensor-product qubit states and falsified the possibility of a decomposition via factorizable rebit states.

    With high statistical significance, we certified the theoretically predicted properties of the states produced.
    First, we showed that the similarities with the targeted two-rebit states are well above $99\%$.
    And a witnessing condition revealed a rebit entanglement close to optimum.
    Furthermore, we reconstructed the optimal decomposition of the mixed-state density operators in terms of tensor-product states, which has previously only applied to almost perfectly pure states \cite{SMBBS19}, such as Bell states which are entangled with respect to both real and complex numbers.
    Thereby, we quantified that the best decomposition of the produced states in terms of product rebit states is close to the maximally possible distance to the actually produced two-rebit state. 

    From a fundamental perspective, our demonstration shows that quantum physics in composite systems does indeed require complex Hilbert spaces for a complete description of the underlying states.
    Conversely, imposing a world with real-valued multipartite wave functions would be incomplete \cite{RTWTTGAN21}.
    On the more practical side, our analysis also demonstrates the importance of nontrivial phases in quantum information protocols that employ entangled states.
    While much attention is devoted to producing entangled states with uniformly distributed amplitudes of Schmidt coefficients, such as Bell states for two qubits, our findings imply that phases between multiple entangled states can be valuable as well.
    Namely, when focusing on real-valued expansions alone, one effectively restricts oneself to an incomplete scenario with real numbers and ignores the additional resources that are offered by a complex-valued description \cite{WKRSXLGS20}.

    Our techniques employed here have been generalized to multipartite entanglement beyond qubits in complex spaces.
    Similar generalizations (for witnessing \cite{SV13} and quasiprobabilities \cite{SW18}) can be adopted in the future when analyzing entanglement in real spaces with multiple subsystems and higher dimensionalities, e.g., as deemed useful in multipartite quantum communications with many partners and large encoding alphabets \cite{W12}.
    Therefore, our studies open a path towards characterizing previously untapped nonlocal quantum coherence properties for applications in quantum technology.

\begin{acknowledgments}
    The Integrated Quantum Optics group acknowledges financial support through the European Commission through the ERC project QuPoPCoRN (Grant No. 725366) and the H2020-FETFLAG-2018-03 project PhoG (Grant No. 820365), as well as funding through the Gottfried Wilhelm Leibniz-Preis (Grant No. SI1115/3-1).
\end{acknowledgments}


\section*{Supplemental Material}
\appendix

\section{Preliminaries}\label{SM:Prelim}

    We begin with introducing the (mostly standard) notation, as used throughout the paper.
    The Pauli operators in terms of the computational basis read
	\begin{eqnarray}
	\begin{aligned}
		\hat\sigma_z=|1\rangle\langle 1|-|0\rangle\langle 0|,
		\quad
		\hat\sigma_x=|0\rangle\langle 1|+|1\rangle\langle 0|,
		\\
		\text{and}\quad
		\hat\sigma_y=i|0\rangle\langle 1|-i|1\rangle\langle 0|.
	\end{aligned}
	\end{eqnarray}
    In addition, the identity on a two-dimensional space is $\hat\sigma_0=\hat1$.
    Eigenvectors of the Pauli matrices are polarization states,
	\begin{subequations}
	\begin{eqnarray}
		|H\rangle&=&|1\rangle=+\hat\sigma_z|H\rangle,
		\\
		|V\rangle&=&|0\rangle=-\hat\sigma_z|V\rangle,
		\\
		|D\rangle&=&\frac{|0\rangle+|1\rangle}{\sqrt 2}=+\hat\sigma_x|D\rangle,
		\\
		|A\rangle&=&\frac{|0\rangle-|1\rangle}{\sqrt 2}=-\hat\sigma_x|A\rangle,
		\\
		|R\rangle&=&\frac{|1\rangle+i|0\rangle}{\sqrt 2}=+\hat\sigma_y|R\rangle,
		\\
		|L\rangle&=&\frac{|1\rangle-i|0\rangle}{\sqrt 2}=-\hat\sigma_y|L\rangle,
	\end{eqnarray}
	\end{subequations}
	which are horizontal, vertical, diagonal, antidiagonal, right-circular, and left-circular polarization, respectively.

    A two-qubit density operator can be expanded in terms of tensor products of Pauli operators,
    \begin{equation}
        \hat\rho=\sum_{\mu,\nu\in\{0,z,x,y\}}\rho_{\mu,\nu}\hat\sigma_\mu\otimes\hat\sigma_\nu.
    \end{equation}
    In particular, the expansion coefficients can be expressed via measured correlations, $\rho_{\mu,\nu}=\langle\hat\sigma_\mu\otimes\hat\sigma_\nu\rangle/4$.
    In general, we can define the correlation matrix
    \begin{equation}
		\label{eq:SMCorrMat}
        \Gamma=\left(\langle\hat\sigma_\mu\otimes\hat\sigma_\nu\rangle\right)_{\mu,\nu\in\{0,z,x,y\}}
    \end{equation}
    that thus contains all information about the quantum state.
    The rebit part of the density operator, $\mathrm{Re}(\hat\rho)$, is then given by a correlation matrix that obeys $\Gamma_{y,\nu}=0=\Gamma_{\mu,y}$ for $\mu,\nu\in\{0,z,x\}$.
    Furthermore, a correlation matrix that represents a state in standard form reads $\Gamma_{\mu,\nu}=\Gamma_{\mu,\mu}\delta_{\mu,\nu}$, where $\delta_{\mu,\nu}=1$ for $\mu=\nu$ and $\delta_{\mu,\nu}=0$ for $\mu\neq\nu$. 

    The Hilbert-Schmidt inner product of two density operators can be written as
    \begin{eqnarray}
        \mathrm{tr}\left(\hat\rho\hat\rho'\right)=\frac{1}{4}\sum_{\mu,\nu\in\{0,z,x,y\}}\Gamma_{\mu,\nu}\Gamma'_{\mu,\nu}
    \end{eqnarray}
    The Hilbert-Schmidt distance is then defined as $\|\hat\rho-\hat\rho'\|=\sqrt{\mathrm{tr}([\hat\rho-\hat\rho']^2)}$.
	And the similarity $S$ between two density matrices is determined through the (Pearson) correlation coefficient,
	\begin{equation}
	    \label{eq:Sim}
	    S=\frac{
	        \mathrm{tr}\left(\hat\rho\hat\rho^\prime\right)
	    }{\sqrt{
	        \mathrm{tr}\left(\hat\rho^{2}\right)\mathrm{tr}\left(\hat\rho^{\prime 2}\right)}
	   }.
	\end{equation}
	This coefficient is one if the two states are identical and zero for a vanishing overlap.

\begin{widetext}
    Finally, for density operators in standard form, the quasiprobabilities $P_\mathrm{std.}$ were computed \cite{SW18}.
    For two rebits, we have
	\begin{equation}
	\begin{aligned}
	    &\left(P_\mathrm{std.}(a,b)\right)_{a,b\in\{H,V,D,A\}}
	    \\
		=&\frac{\Gamma_{0,0}-|\Gamma_{z,z}|-|\Gamma_{x,x}|}{8}
		\begin{pmatrix}
			1 & 1 & 0 & 0
			\\
			1 & 1 & 0 & 0
			\\
			0 & 0 & 1 & 1
			\\
			0 & 0 & 1 & 1
		\end{pmatrix}
		+\frac{1}{4}
		\begin{pmatrix}
			|\Gamma_{z,z}|+\Gamma_{z,z} & |\Gamma_{z,z}|-\Gamma_{z,z} & 0 & 0
			\\
			|\Gamma_{z,z}|-\Gamma_{z,z} & |\Gamma_{z,z}|+\Gamma_{z,z} & 0 & 0
			\\
			0 & 0 & |\Gamma_{x,x}|+\Gamma_{x,x} & |\Gamma_{x,x}|-\Gamma_{x,x}
			\\
			0 & 0 & |\Gamma_{x,x}|-\Gamma_{x,x} & |\Gamma_{x,x}|+\Gamma_{x,x}
		\end{pmatrix}.
	\end{aligned}
	\end{equation}
	And the two-qubit scenario yields (with $Q=\Gamma_{0,0}-|\Gamma_{z,z}|-|\Gamma_{x,x}|-|\Gamma_{y,y}|$)
	\begin{equation}
	\begin{aligned}
	    &\left(P_\mathrm{std.}(a,b)\right)_{a,b\in\{H,V,D,A,R,L\}}
	    \\
		=&\frac{Q}{12}
		\begin{pmatrix}
			1 & 1 & 0 & 0 & 0 & 0
			\\
			1 & 1 & 0 & 0 & 0 & 0
			\\
			0 & 0 & 1 & 1 & 0 & 0
			\\
			0 & 0 & 1 & 1 & 0 & 0
			\\
			0 & 0 & 0 & 0 & 1 & 1
			\\
			0 & 0 & 0 & 0 & 1 & 1
		\end{pmatrix}
		+\frac{1}{4}
		\begin{pmatrix}
			|\Gamma_{z,z}|+\Gamma_{z,z} & |\Gamma_{z,z}|-\Gamma_{z,z} & 0 & 0 & 0 & 0
			\\
			|\Gamma_{z,z}|-\Gamma_{z,z} & |\Gamma_{z,z}|+\Gamma_{z,z} & 0 & 0 & 0 & 0
			\\
			0 & 0 & |\Gamma_{x,x}|+\Gamma_{x,x} & |\Gamma_{x,x}|-\Gamma_{x,x} & 0 & 0
			\\
			0 & 0 & |\Gamma_{x,x}|-\Gamma_{x,x} & |\Gamma_{x,x}|+\Gamma_{x,x} & 0 & 0
			\\
			0 & 0 & 0 & 0 & |\Gamma_{y,y}|+\Gamma_{y,y} & |\Gamma_{y,y}|-\Gamma_{y,y}
			\\
			0 & 0 & 0 & 0 & |\Gamma_{y,y}|-\Gamma_{y,y} & |\Gamma_{y,y}|+\Gamma_{y,y}
		\end{pmatrix}.
	\end{aligned}
	\end{equation}
	In addition, the separable transformation $(\hat A\otimes\hat B)\hat\rho(\hat A\otimes\hat B)^\dag/\mathrm{tr}[(\hat A\otimes\hat B)\hat\rho(\hat A\otimes\hat B)^\dag]$ results in rescaled quasiprobabilities,
	\begin{equation}
	    P(a',b')=\frac{P_\mathrm{std.}(a,b)\langle a|\hat A^\dag\hat A|a\rangle\langle b|\hat B^\dag\hat B|b\rangle}{\sum_{a,b}P(a,b)\langle a|\hat A^\dag\hat A|a\rangle\langle b|\hat B^\dag\hat B|b\rangle}
	\end{equation}
	for transformed states $|a'\rangle=\hat A|a\rangle/\sqrt{\langle a|\hat A^\dag\hat A|a\rangle}$ and $|b'\rangle=\hat B|b\rangle/\sqrt{\langle b|\hat B^\dag\hat B|b\rangle}$.
	Recall that $\dag$ is identical to the transposition $\mathrm{T}$ in the rebit case.
    It was also shown in theory \cite{SW18} that $\sum_{a',b'}P(a',b')|a',b'\rangle\langle a',b'|$ describes the full two-qubit state.
    But in the rebit case, this expansion cannot resolve the $\hat\sigma_y\otimes\hat\sigma_y$ component \cite{SW18}.
\end{widetext}

\section{Witnessing}\label{SM:Witness}

    For deriving our entanglement witnessing approach over real Hilbert spaces, we directly follow the derivation in Ref. \cite{SV09} for complex spaces.
    That is, we determine optimal expectation values through the function $G(a,b)=\langle a,b|\hat L|a,b\rangle$, where $\hat L=\hat L^\mathrm{T}\in\mathbb R^{(d_Ad_B)\times (d_Ad_B)}$, subject to the constraints of normalization, $C_A(a,b)=\langle a|a\rangle-1=0$ and $C_B(a,b)=\langle b|b\rangle-1=0$.
    Note that because of convexity, it suffices to optimize over pure states to obtain the maximal and minimal bounds for mixed separable states.
    Using Lagrangian multipliers $g^{(\mathbb R)}_A$ and $g^{(\mathbb R)}_B$, the optimization yields
    \begin{equation}
    \begin{aligned}
        0&=\nabla_a G
        -g_A^{(\mathbb R)}\nabla_a C_A
        -g_B^{(\mathbb R)}\nabla_a C_B,
        \\
        0&=\nabla_b G
        -g_A^{(\mathbb R)}\nabla_b C_A
        -g_B^{(\mathbb R)}\nabla_b C_B
        \\
        \Leftrightarrow\quad
        0&=2L_b|a\rangle
        -2g_A^{(\mathbb R)}|a\rangle
        -0,
        \\
        0&=L_a|b\rangle
        -0
        -2g_B^{(\mathbb R)}|b\rangle,
    \end{aligned}
    \end{equation}
    applying the notation $(\langle a|\otimes\langle b|)\hat L(|a\rangle\otimes|b\rangle)=\langle a|\hat L_b|a\rangle=\langle b|\hat L_a|b\rangle$ and where $\nabla_{a/b}$ is the gradient with respect to the vector for $a/b$.
    The third and fourth line can be multiplied with $\langle a|$ and $\langle b|$, respectively, proving $g_A^{(\mathbb R)}=\langle a,b|\hat L|a,b\rangle=g_B^{(\mathbb R)}$.
    This allows us to drop the subscript, $g^{(\mathbb R)}=g_A^{(\mathbb R)}=g_B^{(\mathbb R)}$, and conclude that the Lagrangian multiplier corresponds to an optimal expectation value of $\hat L$.

    Therefore, the global maximum $g^{(\mathbb R)}_{\max}$ and global minimum $g^{(\mathbb R)}_{\min}$ are obtained as the maximum and minimum over all optimal expectations values $g^{(\mathbb R)}$.
    From the optimization, we also directly obtain the separability eigenvalue equations, $\hat L_b|a\rangle=g^{(\mathbb R)}|a\rangle$ and $\hat L_a|b\rangle=g^{(\mathbb R)}|b\rangle$ (third and fourth line in the above equation), as presented in the main text.

    For completeness, let us also mention that the ordinary eigenvalue equation reads $\hat L|\psi\rangle=g|\psi\rangle$, and that the complex separability eigenvalue equations takes the same form as the real ones derived above, however, allowing for $|a\rangle\in\mathbb C^{d_A}$ and $|b\rangle\in\mathbb C^{d_B}$ \cite{SV09}.

    As an example, we consider a family of operators of the form
    \begin{equation}
        \hat L
        =L_{z}\hat\sigma_z\otimes\hat\sigma_z
        +L_{x}\hat\sigma_x\otimes\hat\sigma_x
        +L_{y}\hat\sigma_y\otimes\hat\sigma_y.
    \end{equation}
    These observables enable us to consider all witnesses that are discussed and applied in the main text---by setting either $L_y=0$ or $L_z=L_x=0$.
    It it worth mentioning that this defines a standard form of general operators $\sum_{\mu,\nu\in\{z,x,y\}}L'_{\mu,\nu}\hat\sigma_\mu\otimes\hat\sigma_\nu$ when applying local unitary transformations, which rotate separability eigenvectors accordingly but do not alter separability eigenvalues \cite{SV09}.

    For convenience, we expand Alice's normalized state as $|a\rangle\langle a|=(\hat\sigma_0+\alpha_z\hat\sigma_z+\alpha_x\hat\sigma_x+\alpha_y\hat\sigma_y)$, with $\alpha_y=0$ for real Hilbert spaces and $\sum_{\mu}\alpha_\mu^2=1$ so that the expansions describe proper pure states.
    Now, the reduced operator for Bob's system reads $\hat L_a=\sum_{\mu}L_\mu\alpha_\mu\hat\sigma_\mu$.
    The eigenvectors to this operator \cite{CommentEigenvectors} expand as $|b\rangle\langle b|=(\hat\sigma_0\pm_B\sum_\mu L_\mu\alpha_\mu\hat\sigma_\mu/B)/2$, where $B=[\sum_\mu L_\mu^2\alpha_\mu^2]^{1/2}$.
    Thus, for the second separability eigenvalue equation, we obtain the reduced operator $\hat L_b=\pm_B\sum_\mu L_\mu^2\alpha_\mu\hat\sigma_\mu/B$.
    This operator has the eigenvectors $(\hat\sigma_0\pm_A\sum_\mu L_\mu^2\alpha_\mu/A)/2$, where $A=[\sum_\mu L_\mu^4\alpha_\mu^2]^{1/2}$ \cite{CommentEigenvectors}.
    These solutions are identical to $|a\rangle\langle a|$ if $\alpha_\mu=\pm_AL_{\mu}^2\alpha_\mu/A$ is satisfied for all $\mu$.
    In the general case, where $L_\mu^2\neq L_\nu^2$ for $\mu\neq\nu$, this results in only one expansion coefficient $\alpha_\mu$ that is not equal to zero, i.e., $|a\rangle\langle a|=(\hat\sigma_0\pm_A\hat\sigma_\mu)/2$.
    This then gives $|b\rangle\langle b|=(\hat\sigma_0\pm_B\hat\sigma_\mu)/2$.
    The separability eigenvalue is given by the expectation value as derived above, $\mathrm{tr}(\hat L|a,b\rangle\langle a,b|)=\pm_A\pm_B L_\mu$.
    Since our equations are polynomial ones in $|a\rangle$ and $|b\rangle$, thus continuous, we get the same solutions even when $L_\mu^2\to L_\nu^2$.

    In conclusions, we find for the family of operators under study the real separability eigenvectors and their separability eigenvalues as
    \begin{equation}
    \begin{aligned}
        |ab\rangle\in&\{
            |HH\rangle,|HV\rangle,|VH\rangle,|VV\rangle,
            |DD\rangle,|DA\rangle,|AD\rangle,|AA\rangle
        \},
        \\
        g^{(\mathbb R)}\in&\{
            L_z,-L_z,-L_z,L_z,
            L_x,-L_x,-L_x,L_x,
        \}.
    \end{aligned}
    \end{equation}
    In addition, we obtain $|ab\rangle\in\{|RR\rangle,|RL\rangle,|LR\rangle,|LL\rangle\}$ for $g^{(\mathbb C)}\in\{L_y,-L_y,-L_y,L_y\}$ for the two-qubit case, together with all aforementioned solutions for real Hilbert spaces.
    For completeness, the ordinary eigenvectors are Bell states,
    \begin{equation}
    \begin{aligned}
        |\psi\rangle\in&\left\{
            \frac{|HH\rangle+|VV\rangle}{\sqrt{2}},
            \frac{|HH\rangle-|VV\rangle}{\sqrt{2}},
        \right.
        \\
        &\left.
            \frac{|HV\rangle+|VH\rangle}{\sqrt{2}},
            \frac{|HV\rangle-|VH\rangle}{\sqrt{2}}
        \right\},
    \end{aligned}
    \end{equation}
    for ordinary eigenvalues $g\in\{L_z+L_x-L_y,L_z-L_x+L_y,-L_z+L_x+L_y,-L_z-L_x-L_y\}$.

\section{Data processing}\label{SM:Data}

    Our data processing closely follows the approach that was formulated in the Supplemental Material of Ref. \cite{SMBBS19}, including the utilized error estimations.
    For self-consistency, we briefly recapitulate those methods here.
    Specifically, we formulate the approach in terms of directly accessible correlations $\langle\hat\sigma_\mu\otimes\hat\sigma_\nu\rangle$.

\begin{figure}[t]
    \includegraphics[width=\columnwidth]{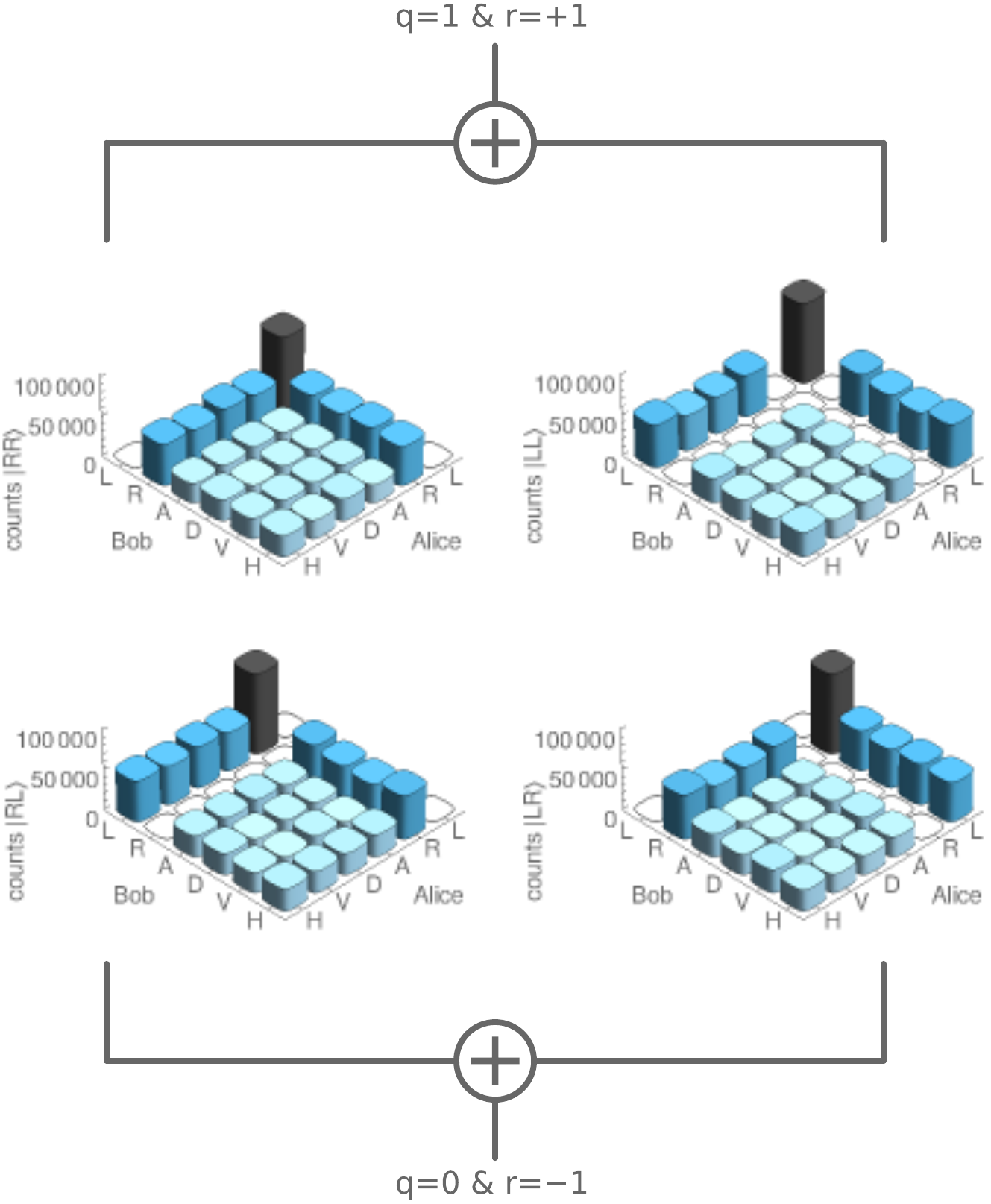}
    \caption{%
        Raw data from circularly polarized states (top-left: $|RR\rangle$; top-right: $|LL\rangle$; bottom-left: $|RL\rangle$; bottom-right: $|LR\rangle$).
        The datasets in the top and bottom row are combined for obtaining the CFR states $\propto |LL\rangle\langle LL|+|RR\rangle\langle RR|\propto \hat\sigma_0\otimes\hat\sigma_0+\hat\sigma_y\otimes\hat\sigma_y$ and $\propto |RL\rangle\langle RL|+|LR\rangle\langle LR|\propto \hat\sigma_0\otimes\hat\sigma_0-\hat\sigma_y\otimes\hat\sigma_y$, respectively.
    }
    \label{fig:RawData}
\end{figure}

    Measured data are shown in Fig. \ref{fig:RawData}, and combined as depicted, resulting in the sought-after CFR states.
    As described in Ref. \cite{SMBBS19}, the data allow one to directly determine the experimentally measured correlation matrix [Eq. \eqref{eq:SMCorrMat}]
    \begin{equation}
        \Gamma^\mathrm{(exp)}=\begin{pmatrix}
            \Gamma^\mathrm{(exp)}_{0,0} & \Gamma^\mathrm{(exp)}_{0,z} & \Gamma^\mathrm{(exp)}_{0,x} & \Gamma^\mathrm{(exp)}_{0,y}
            \\
            \Gamma^\mathrm{(exp)}_{z,0} & \Gamma^\mathrm{(exp)}_{z,z} & \Gamma^\mathrm{(exp)}_{z,x} & \Gamma^\mathrm{(exp)}_{z,y}
            \\
            \Gamma^\mathrm{(exp)}_{x,0} & \Gamma^\mathrm{(exp)}_{x,z} & \Gamma^\mathrm{(exp)}_{x,x} & \Gamma^\mathrm{(exp)}_{x,y}
            \\
            \Gamma^\mathrm{(exp)}_{y,0} & \Gamma^\mathrm{(exp)}_{y,z} & \Gamma^\mathrm{(exp)}_{y,x} & \Gamma^\mathrm{(exp)}_{y,y}
        \end{pmatrix},
    \end{equation}
    together with the standard deviation $\sigma(\Gamma^\mathrm{(exp)}_{\mu,\nu})$ for each component, $\mu,\nu\in\{0,z,x,y\}$.
    As described in Sec. \ref{SM:Prelim}, this enables us to reconstruct the density operator,
    \begin{equation}
        \hat\rho^\mathrm{(exp)}=\frac{1}{4}\sum_{\mu,\nu\in\{0,z,x,y\}}\Gamma^\mathrm{(exp)}_{\mu,\nu}\hat\sigma_\mu\otimes\hat\sigma_\nu.
    \end{equation}
    And the similarity $S$ [Eq. \eqref{eq:Sim}] with the targeted CFR state, which is represented by a correlation matrix $\Gamma^\mathrm{(tar)}=\mathrm{diag}(1,0,0,\pm1)$ for $q\in\{1,0\}$, can be computed via
    \begin{equation}
        S=\frac{\sum\limits_{\mu,\nu\in\{0,z,x,y\}}\Gamma^\mathrm{(exp)}_{\mu,\nu}\Gamma^\mathrm{(tar)}_{\mu,\nu}}{\sqrt{
            \left(\sum\limits_{\mu,\nu\in\{0,z,x,y\}}\Gamma^{\mathrm{(exp)}2}_{\mu,\nu}\right)
            \left(\sum\limits_{\mu,\nu\in\{0,z,x,y\}}\Gamma^{\mathrm{(tar)}2}_{\mu,\nu}\right)
        }},
    \end{equation}
    where a quadratic error propagation is used to determine uncertainties, resulting in the quantities presented in the main text.
    In addition, the expectation value of the witness $\hat L=\hat\sigma_y\otimes\hat\sigma_y$ is simply given by the matrix element $\Gamma^\mathrm{(exp)}_{y,y}$, together with its uncertainty.

    Determining the standard form for the complex case is done in two steps \cite{SMBBS19}.
    In the first step, local transformations $A_1\in\mathbb R^{4\times 4}$ and $B_1\in\mathbb R^{4\times 4}$ are constructed to obtain
    \begin{equation}
    \begin{aligned}
        \Gamma^\mathrm{(aux)} = {}& A_1\Gamma^\mathrm{(exp)}B_1^\mathrm{T}
        \\
        = {}& \begin{pmatrix}
            \Gamma^\mathrm{(aux)}_{0,0} & 0 & 0 & 0
            \\
            0 & \Gamma^\mathrm{(aux)}_{z,z} & \Gamma^\mathrm{(aux)}_{z,x} & \Gamma^\mathrm{(aux)}_{z,y}
            \\
            0 & \Gamma^\mathrm{(aux)}_{x,z} & \Gamma^\mathrm{(aux)}_{x,x} & \Gamma^\mathrm{(aux)}_{x,y}
            \\
            0 & \Gamma^\mathrm{(aux)}_{y,z} & \Gamma^\mathrm{(aux)}_{y,x} & \Gamma^\mathrm{(aux)}_{y,y}
        \end{pmatrix}.
    \end{aligned}
    \end{equation}
    The second step then yields the standard form,
    \begin{equation}
    \begin{aligned}
        \Gamma^\mathrm{(std)} = {}& A_2\Gamma^\mathrm{(aux)}B_2^\mathrm{T}
        \\
        = {}& \begin{pmatrix}
            \Gamma^\mathrm{(std)}_{0,0} & 0 & 0 & 0
            \\
            0 & \Gamma^\mathrm{(std)}_{z,z} & 0 & 0
            \\
            0 & 0 & \Gamma^\mathrm{(std)}_{x,x} & 0
            \\
            0 & 0 & 0 & \Gamma^\mathrm{(std)}_{y,y}
        \end{pmatrix},
    \end{aligned}
    \end{equation}
    via rotations $A_2,B_2\in\mathbb R^4$.
    Thus, the combination of both results in local maps for Alice and Bob, $A=A_2A_1$ and $B=B_2B_1$, such that
    \begin{equation}
        \Gamma^\mathrm{(std)}=A\Gamma^\mathrm{(exp)} B^\mathrm{T}
    \end{equation}
    yields the diagonal standard form.
    In terms of density operators, this corresponds to $\hat\rho_\mathrm{std.}\propto (\hat A\otimes \hat B)^{-1}\hat\rho(\hat A^\dag\otimes \hat B^\dag)^{-1}$.
    Furthermore, we can now apply the formula for qubit quasiprobabilities $P_\mathrm{std.}$ to the standard form, as given in Sec. \ref{SM:Prelim}.

    The expansion in terms of quasiprobabilities and factorizable states for the standard form is given via $|a\rangle\langle a|$ and $|b\rangle\langle b|$ that correspond to the eigenstates $|w\rangle$ of the Pauli operators, which can be also represented by $\Gamma_w=(\langle\hat\sigma_\mu\rangle)_{\mu\in\{0,z,x,y\}}$,
    \begin{equation}
        \Gamma_w\in\left\{
            \begin{pmatrix}1\\1\\0\\0\end{pmatrix},
            \begin{pmatrix}1\\-1\\0\\0\end{pmatrix},
            \begin{pmatrix}1\\0\\1\\0\end{pmatrix},
            \begin{pmatrix}1\\0\\-1\\0\end{pmatrix},
            \begin{pmatrix}1\\0\\0\\1\end{pmatrix},
            \begin{pmatrix}1\\0\\0\\-1\end{pmatrix}
        \right\},
    \end{equation}
    with $w\in\{H,V,D,A,R,L\}$.
    In particular, this yields $\Gamma^\mathrm{(std)}=\sum_{a,b}P_\mathrm{std}(a,b)\Gamma_a{\Gamma_b}^\mathrm{T}$.
    To get the distribution of the experimentally obtained correlation matrix, we use the above transformations, $\Gamma^\mathrm{(exp)}=\sum_{a,b}P_\mathrm{std}(a,b)A^{-1}\Gamma_a{\Gamma_b}^\mathrm{T}B^{-\mathrm{T}}$, as described in Sec. \ref{SM:Prelim} for the density operator.
    Here, we thereby obtain normalized $\Gamma_{a'}=A^{-1}\Gamma_a/(A^{-1}\Gamma_a)_0$ and $\Gamma_{b'}=B^{-1}\Gamma_b/(B^{-1}\Gamma_b)_0$ to represent the local states, where $(\cdots)_j$ generally denotes the $j$th vector entry for $j\in\{0,z,x,y\}$, and $P(a',b')=P_\mathrm{std}(a,b)(A^{-1}\Gamma_a)_0(B^{-1}\Gamma_b)_0$ as their quasiprobabilities, which is depicted in the main part of the paper.

    Then, the resulting local expansion is given by
    \begin{equation}
        \Gamma^\mathrm{(loc)}=\sum_{a',b'}P(a',b')\Gamma_{a'}\Gamma_{b'}^\mathrm{T},
    \end{equation}
    which defines the state $\hat\rho^\mathrm{(loc)}=\sum_{\mu,\nu\in\{0,z,x,y\}}\Gamma^\mathrm{(loc)}_{\mu,\nu}\hat\sigma_\mu\otimes\hat\sigma_\nu/4$.
    We can now compute the Hilbert-Schmidt distance to the initially reconstructed state $\hat\rho^\mathrm{(exp)}$,
    \begin{equation}
        \|\hat\rho^\mathrm{(exp)}-\hat\rho^\mathrm{(loc)}\|
        =\sqrt{\frac{1}{4}\sum_{\mu,\nu}\left(
            \Gamma^\mathrm{(exp)}_{\mu,\nu}-\Gamma^\mathrm{(loc)}_{\mu,\nu}
        \right)^2}.
    \end{equation}
    The theoretical prediction is that this distance is zero for complex Hilbert spaces.

    Errors are propagated using the following, standard Monte Carlo approach:
    A sufficiently large sample of $10\,000$ correlation matrices that correspond to physical states are produced with entries that are normally distributed with means $\Gamma^\mathrm{(exp)}_{\mu,\nu}$ and standard deviations $\sigma(\Gamma^\mathrm{(exp)}_{\mu,\nu})$.
    All aforementioned quantities and decompositions are computed for each sample element as described above and the standard deviation of the results then estimates the propagated uncertainties.

    For the rebit scenario, we begin our treatment with the real part of the density operator.
    In terms of correlation matrices, this means that we replace $\Gamma^\mathrm{(exp)}$ in the algorithm with
    \begin{equation}
        \Gamma^{(\mathbb R)}=\begin{pmatrix}
            \Gamma^\mathrm{(exp)}_{0,0} & \Gamma^\mathrm{(exp)}_{0,z} & \Gamma^\mathrm{(exp)}_{0,x} & 0
            \\
            \Gamma^\mathrm{(exp)}_{z,0} & \Gamma^\mathrm{(exp)}_{z,z} & \Gamma^\mathrm{(exp)}_{z,x} & 0
            \\
            \Gamma^\mathrm{(exp)}_{x,0} & \Gamma^\mathrm{(exp)}_{x,z} & \Gamma^\mathrm{(exp)}_{x,x} & 0
            \\
            0 & 0 & 0 & \Gamma^\mathrm{(exp)}_{y,y}
        \end{pmatrix}
    \end{equation}
    and use $P_\mathrm{std.}$ for rebits from Sec. \ref{SM:Prelim}.
    Of course, we still determine the distance of the locally reconstructed state, i.e., $\Gamma^\mathrm{(loc)}$, to the actually produced one, represented by the full matrix $\Gamma^\mathrm{(exp)}$.
    It is worth pointing out that $\Gamma^{(\mathbb R)}$ is already diagonalized with respect to the $y$ component.
    Hence, by construction \cite{SMBBS19}, the local transformation matrices $A$ and $B$ leave $y$ entries invariant in the rebit scenario.



\begin{thebibliography}{99}
\bibitem{CommentEigenvectors}
		The two eigenvectors $|\psi_\pm\rangle$ of a matrix $\hat M=W\hat\sigma_0+Z\hat\sigma_z+X\hat\sigma_x+Y\hat\sigma_y$ can be written via the dyadic product as $|\psi_\pm\rangle\langle \psi_\pm|=(\hat\sigma_0\pm R^{-1}[Z\hat\sigma_z+X\hat\sigma_x+Y\hat\sigma_y])/2$, where $R=\sqrt{Z^2+X^2+Y^2}$.
\end{thebibliography}
\end{document}